\documentclass[pra,twocolumn,superscriptaddress,floatfix,showpacs]{revtex4}
\usepackage{amsmath,amsfonts,epsf,dcolumn,natbib,graphicx}
\usepackage{ascii}
\usepackage{xcolor}
\newcommand{\half}{\tfrac{1}{2}}
\newcommand{\BIG}[2]{\mbox{$\left#1\vbox to #2{}\right. $}}
\newcommand{\be}{\begin{equation}}
\newcommand{\ee}{\end{equation}}
\newcommand{\bea}{\begin{eqnarray}}
\newcommand{\eea}{\end{eqnarray}}

\begin{document}
\bibliographystyle{plainnat}
\title{Finite-temperature orbital-free DFT 
molecular dynamics: \\ coupling {\sc Profess} and {\sc Quantum Espresso}}

\author{Valentin V.~Karasiev}
\email{vkarasev@qtp.ufl.edu}
\affiliation{Quantum Theory Project, 
Departments of Physics and of Chemistry, P.O. Box 118435, 
University of Florida, Gainesville FL 32611-8435}
\author{Travis Sjostrom}
\affiliation{Theoretical Division, Los Alamos National Laboratory, Los Alamos, NM 87545}
\author{S.B.~Trickey}
\affiliation{Quantum Theory Project, 
Departments of Physics and of Chemistry, P.O. Box 118435, 
University of Florida, Gainesville FL 32611-8435}

\date{03 June 2014}

\begin{abstract}

Implementation of orbital-free free-energy functionals in the {\sc Profess} code
and the coupling of {\sc Profess} with the {\sc Quantum Espresso} code are 
described.  The combination enables orbital-free DFT to drive 
{\it ab initio} molecular dynamics simulations on the same footing 
(algorithms, thermostats, convergence parameters, etc.) as
for Kohn-Sham (KS) DFT.  All the non-interacting free-energy functionals implemented
are single-point: the local density approximation (LDA; also known as 
finite-T Thomas-Fermi, ftTF),  the second-order gradient approximation 
(SGA or finite-T gradient-corrected TF), and our recently introduced 
finite-T generalized gradient approximations (ftGGA). 
Elimination of the KS orbital bottleneck via orbital-free
methodology enables high-T simulations on ordinary computers,
whereas those simulations would be costly or even 
prohibitively time-consuming for KS molecular dynamics (MD) on very high-performance 
computer systems.  
Example MD simulations on H 
over a temperature range $2,000$ K $ \le \mathrm{T} \le 4,000,000$ K are reported, 
with timings on small clusters (16-128 cores) and even laptops.  With
respect to KS-driven calculations, the orbital-free calculations are between 
a few times through a few hundreds of times faster.  

\end{abstract} 


\maketitle

\renewcommand{\baselinestretch}{1.05}\rm

\section{Introduction}

Orbital-free density functional theory (OF-DFT) in principle provides
an exact quantum-mechanical description of many-electron systems,
both in the ground-state and at non-zero temperature T.  
Computationally, OF-DFT should be drastically less expensive than
conventional Kohn-Sham (KS) DFT \cite{KohnSham} at all T. 
The reason is well-known.  Explicit use of the KS orbitals involves
orthogonalization, which causes a computational cost scaling no better than  
$N_b^3$, with $N_b$ the number of occupied KS energy levels.  At T=0 K,
$N_b$ is proportional to the number of electrons in the system, hence
grows with system size and complexity.  Non-zero T makes matters worse,
as the Fermi-Dirac distribution increases the number of computationally 
significant (compared to machine precision) occupation numbers 
relative to the ground state.  
Thus, the use of 
KS-DFT to drive {\it ab  initio} molecular dynamics (AIMD)
\cite{Barnett93,MarxHutter2000,Tse2002,MarxHutter2009,Kuhne12} is
circumscribed by computer resource limits
\cite{CollinsCECAM2012,GerickePNP14} because a KS-DFT calculation
must be done at each MD step.   This scaling behavior is a notable   
challenge to detailed computational exploration of warm dense matter (WDM), 
which has temperatures of hundreds of kK at 
material densities from near ambient to several-fold compression. 
In contrast, the OF-DFT 
computational cost should scale essentially linearly with system size,
irrespective of T.  

Nevertheless, the KS decomposition is the appropriate framework for
formulating OF-DFT as a useful computational tool for
several reasons. Both the exchange (X) energy and the kinetic energy (KE)
contribution to the DFT correlation (C) energy are defined in terms
of the KS decomposition.  And nearly 50 years of development of
effective approximate XC functionals has taken place in that framework. 
The main ingredient, therefore, of ground-state OF-DFT is the 
non-interacting (or KS) KE functional.  
For non-zero T, the corresponding ingredient is the non-interacting 
free-energy functional, with contributions from the non-interacting 
KE and non-interacting entropy. Both ground-state and finite-T also 
require an orbital-free XC functional, of course.

Though the OF-DFT Euler equation can be cast quite easily into
the form of a one-orbital (proportional to the square root of
the density) KS equation with an extra potential, use of a 
standard KS code to solve that equation is not a viable 
strategy \cite{ChanCohenHandy2001,KarasievTrickeyCPC2012}.  Direct
minimization techniques are required.   
{\sc Profess} \cite{Ho..Carter08,Hung..Carter10} is the only 
widely distributed OF-DFT computational package of which we are
aware \cite{AbInitandNWChem} which provides such techniques for 
computation of the ground state energy, electron density, inter-atomic
electronic forces, and stress tensor with both periodic and Dirichlet 
boundary conditions (PBCs and DBCs). {\sc Profess} may be used at 
a single ionic geometry 
(``single-point calculations''), wherein the total energy is minimized
with respect to the electron density by one of the direct optimization methods
implemented in the code, such as nonlinear conjugate gradient (CG) 
minimization and the truncated Newton (TN) method 
(see \cite{Ho..Carter08} for details and references) or for 
geometry optimization to find energetically optimal 
ion positions and cell vectors.  

{\sc Profess} is designed, however, as a ground-state OF-DFT code
for optimization of two-point KE functionals
containing a non-local part, 
\be
T_{\mathrm{NL}}[n] = \int\, d{\mathbf r}d{\mathbf r}^\prime 
n^\lambda({\mathbf r}) K_{\mathrm{s}}[n({\mathbf r}),%
n({\mathbf r}^\prime),{\mathbf r},{\mathbf r}^\prime] %
n^\gamma({\mathbf r}^\prime)\,.
\label{twopoint}
\ee
Here 
$n({\mathbf r})$ is the electron number density,  
$\lambda + \gamma = 8/3$,  and the dimensionless kernel 
$K_{\mathrm{s}}[n({\mathbf r}),n({\mathbf r}^\prime),{\mathbf r} %
,{\mathbf r}^\prime]$  is a type of response function.  
Commonly the non-local form Eq.\ (\ref{twopoint}) is used in conjunction
with the Thomas-Fermi (TF) \cite{Thomas,Fermi}, and 
von Weizs\"acker \cite{Weizsacker} functionals to give the approximate KS KE 
functional,  $T_{\mathrm{s}}$:
\bea
T_{\mathrm s} &\approx& T_{\mathrm{TF}} + T_{\mathrm W} + T_{\mathrm{NL}} \label{TsTtfTvwTnl} \\
T_{\mathrm{TF}}[n] &=& \int d{\mathbf r} \tau_0^{\mathrm{TF}}(n) \label{TF_zeroT} \\
\tau_0^{\mathrm{TF}}(n) &:=& \frac{3}{10}(3\pi^2)^{2/3}n^{5/3} %
\equiv c_{\mathrm{TF}}n^{5/3} \label{tauTFdefn}  \\
T_{\mathrm W}[n] &:=& 
\frac{1}{8}\int d{\mathbf r}  \frac{|\nabla n({\mathbf r})|^2}
 {n({\mathbf r})}   \equiv \int  d{\mathbf r}  t_0^{\mathrm W} ([n];{\mathbf r}) %
\label{T_Wdefn}
\eea
(We use Hartree a.u. unless explicitly
noted to the contrary.  1 E$_H$ = 27.2116 eV, 1 bohr = 0.529177 \AA.)
The 
literature of such two-point 
approximations is accessible through 
Refs.\ \onlinecite{Ho..Carter08,Hung..Carter10} as well as the 
earlier review article by Wang and Carter \cite{WangCarter2000}.  

In the interest of computational speed for AIMD, our work has emphasized 
one-point functionals.  As distributed, the only one-point functionals 
included in {\sc Profess} are $T_{\mathrm{TF}}$ and an 
empirically parametrized linear combination,
\be
T_{\mathrm{TFvW}\{\lambda,\mu\}} =\lambda T_{\mathrm{TF}} + \mu T_{\mathrm{W}} \; , \; %
\label{TTFvW} \\
\ee
with either the Thomas-Fermi ($\lambda=1$, $\mu=0$) or the von Weizs\"acker 
($\lambda=0$, $\mu=1$) term taken as the starting point and the other 
term assumed to be its correction.

Implementation of more refined one-point 
functionals in {\sc Profess}, specifically our earlier    
T = 0 K generalized gradient approximation (GGA) kinetic energy functionals
\cite{Perspectives,PRB80},  was reported without detail 
by two of us \cite{KarasievTrickeyCPC2012}.  
In addition to those earlier T = 0 K modifications, the  
enhanced version of {\sc Profess} presented as a major part of
this work includes our 
recently published GGA non-interacting free-energy
functionals \cite{KarasievSjostromTrickey12A,VT84F} and  a new 
explicitly T-dependent XC functional \cite{LSDA-PIMC} as well as 
two earlier ones \cite{PDW84,PDW2000}.

\begin{figure}
\includegraphics*[angle=-00,height=10.cm]{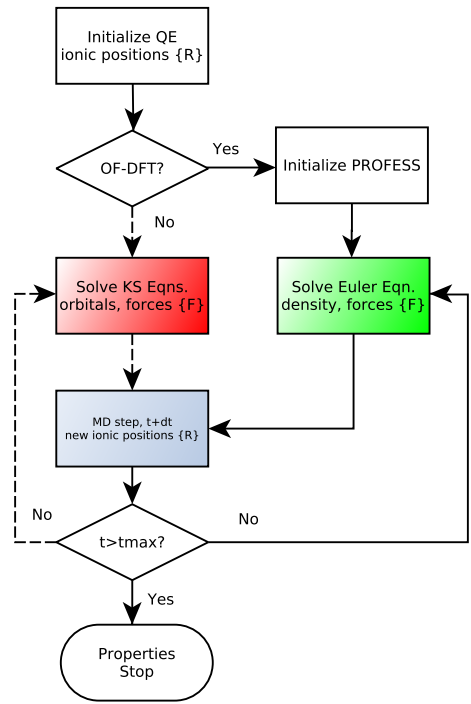}
\caption{
Flow chart for AIMD simulation with {\sc Profess}$@${\sc Q-Espresso} 
package.  Here $dt$ is the MD time step and $t_{\mathrm{max}}$ is 
the total simulation time.
}
\label{P-QE}
\end{figure}

All of these enhancements are in the context of coupling the modified
{\sc Profess} code to a fully-featured, freely available KS-AIMD code,
{\sc Quantum Espresso} \cite{QEspresso}.  The package of
modifications and interfacing, called {\sc Profess}@{\sc Q-Espresso}, 
provides a new, finite-T OF-DFT-AIMD capability useful from the
ground state to far into the WDM regime.  Fig.\ \ref{P-QE} depicts
the relationship between the two codes and the flow of calculation
enabled by the new interface and associated modifications. Detailed
commentary is below.

Motivation for the combination is straight-forward.  Whether the 
AIMD is KS-based
or OF-DFT-based, the combination means that the same algorithms are used 
to execute the MD and to thermostat it in the NVT ensemble.  
For functional development work,    
this artifact-free uniformity of treatment is 
important.  For materials simulation
research generally, the {\sc Profess}@{\sc Q-Espresso} package
makes up-to-date OF-DFT functionals and optimization techniques
available to those who presently run KS-AIMD calculations, again on
a uniform, artifact-free basis \cite{OurFrontiersInWDM.2014}.  Third, use
of packages such as {\sc Profess} and {\sc Quantum Espresso} which 
have published source code avoids 
wasteful duplication of software development effort.  

The remainder of the presentation is organized as follows.  
Essentials (for this discussion) of 
finite-T OF-DFT are described in the 
next Section. Section III gives expressions for 
the ground state KE and finite-T non-interacting free-energy functionals
implemented in the code, along  with their 
functional derivatives and stress tensor components.  Sect.\ IV provides
the corresponding information for the XC functionals.   
Section V describes implementation issues, including the local model 
potentials required in  OF-DFT, the patches needed, and the interface 
itself.  Sect.\ VI  gives example  
timings for OF-DFT static and AIMD calculations on Hydrogen. 

\section{Finite-T OF-DFT Essentials}

For fixed ionic positions $\lbrace {\mathbf R} \rbrace$, the grand canonical 
potential of a system with average electron number $N$ at temperature 
$\rm T$ and chemical potential $\mu$  may be expressed as the
density functional \cite{Mermin65,Stoitsov88}
\be 
\Omega[n]= {\mathcal F}[n] + \int d{\mathbf r} 
n({\mathbf r}) \,\lbrace v_{\mathrm{ext}}({\mathbf r})-\mu \rbrace   %
+ E_{\rm{ion}}( \{{\mathbf R}\}) \,,
\label{E1}
\ee
where $v_{\mathrm{ext}}({\mathbf r})$ is the external potential
and $E_{\rm{ion}}(\lbrace {\mathbf R}\rbrace) $ is the ion-ion Coulomb
repulsion energy.  
(For notational simplicity, explicit T-dependence is suppressed except
where necessary.)  
The universal free-energy functional ${\mathcal F}[n]$
\be
{\mathcal F}[n]={\mathcal F}_{\mathrm{s}}[n]+{\mathcal F}_{\mathrm H}[n] + %
{\mathcal F}_{\mathrm xc}[n]\, .
\label{E3}
\ee
has the conventional KS decomposition
into the non-interacting free energy ${\mathcal F}_{\mathrm{s}}[n]$,
(which has both kinetic and entropic contributions)
\be
{\mathcal F}_{\mathrm{s}}[n]={\mathcal T}_{\mathrm{s}}[n]-{\rm T}%
{\mathcal S}_{\mathrm{s}}[n] \, ,
\label{E4}
\ee
the classical Coulomb (or Hartree) interaction energy,  
\be
{\mathcal F}_{\mathrm H}[n]=\half  
\int\int d{\mathbf r}d{\mathbf r}^\prime %
 \frac{n({\mathbf r})n({\mathbf r}^\prime)}{|{\mathbf r}-{\mathbf r}^\prime|} \; .
\label{E2}
\ee
The remainder, namely the difference between the 
interacting and non-interacting 
free-energy components plus the difference between the full 
quantum mechanical electron-electron interaction energy,
${\mathcal U}_{\rm ee}[n]$, and its 
classical part, ${\mathcal F}_{\rm H}[n]$, is the exchange-correlation (XC)
free energy functional
\bea
{\mathcal F}_{\mathrm xc}[n]&\equiv& ({\mathcal T}[n]- %
{\mathcal T}_{\mathrm{s}}[n]) - {\mathrm T}({\mathcal S}[n]- %
{\mathcal S}_{\mathrm{s}}[n]) \nonumber \\
&&+  ({\mathcal U}_{\mathrm ee}[n]-{\mathcal F}_{\mathrm H}[n])\,  .
\label{E5}
\eea

By stipulation, the non-interacting (KS) system must deliver the
same density as the interacting system, whence one has 
the KS effective potential $v_{\mathrm KS}$ and the 
system of coupled one-electron differential equations
for the KS orbitals,
\be
\left\{-\half \nabla^2 +
v_{\rm KS}([n];{\mathbf r})\right\}\varphi_j({\mathbf r})= %
\varepsilon_j \phi_j({\mathbf r}) \; 
\label{E6}
\ee
for the variational optimization.  
The KS potential is the sum of the external, $v_{\mathrm{ext}}$,
Hartree, $v_{\rm H}:=\delta {\mathcal F}_{\rm H}[n]/\delta n$, and
exchange-correlation, 
$v_{\mathrm xc}:=\delta {\mathcal F}_{\rm xc}[n]/\delta n$ contributions.
In terms of the KS orbitals and Fermi-Dirac occupation numbers, the exact 
non-interacting KE and entropy are 
\bea
{\mathcal T}_{\mathrm{s}}[n]&=& \half 
\sum_{j=1}^{\infty} f_j \int d{\mathbf r} \mid \nabla \varphi_j({\mathbf r})%
\mid^2 \label{TorbDefn} \\
{\mathcal S}_{\mathrm{s}}[n]&=&-k_{\mathrm B}
\sum_{j=1}^{\infty}\{f_j\ln f_j+(1-f_j)\ln(1-f_j)\}\;.  
\label{E7}
\eea
The Fermi-Dirac occupation numbers are
\be
f_j \equiv f(\varepsilon_j-\mu):=1/[1+\exp(\beta(\varepsilon_j-\mu))]\,,
\label{E8}
\ee
$\beta:=1/k_B{\rm T}$, the electron number density is 
\be
n({\mathbf r})=\sum_{j=1}^{\infty} f_j |\varphi_j({\mathbf r})|^2\, ,
\label{E9}
\ee
and the average number of electrons is 
$N_e = \int d{\mathbf r} n({\mathbf r})$. 

The orbital-free alternative has 
non-interacting free-energy and entropy functionals, 
\bea
{\mathcal T}_{\mathrm{s}}[n]&=& \int d{\mathbf r} \, \tau_{\mathrm s}[n] \label{tausubs} \\
{\mathcal S}_{\mathrm{s}}[n]&=&  \int d{\mathbf r} \, \sigma_{\mathrm s}[n] \label{sigmasubs} \;,
\eea
which depend explicitly upon the electron density (and its gradients,
Laplacians, etc.) without explicit reference to the KS orbital manifold
$\lbrace \varphi \rbrace$. 
Just as with the XC free energy functional, the exact form of those two
functionals is unknown in general, so approximations must be constructed.
Assuming that one has an orbital-free XC approximation as well,   
minimization of the functional Eq.\ (\ref{E1}) gives the single 
Euler equation
\be
\frac{\delta {\mathcal F}_{\mathrm s}}{\delta n({\mathbf r})} 
=  \mu - v_{\mathrm KS}([n];{\mathbf r})\equiv \mu([n];{\bf r}) \, .
\label{E10}
\ee
The XC contribution appears in $v_{\mathrm KS}$.  

As noted already, solution of the OF-DFT problem is by 
direct minimization of the total electronic free energy
\begin{align}
  \mathcal{F}_{\mathrm{tot}}[n] = \mathcal{F}[n]+\int n(\mathbf{r}) v_{\mathrm{ext}}(\mathbf{r}) d\mathbf{r},
\end{align}
with the constraint that the integral of the density is constant. We use 
the nonlinear conjugate gradient techniques  implemented in  
{\sc Profess}. In contrast, the KS equation must be diagonalized 
in some basis, plane waves in the case of {\sc Quantum Espresso}.  
Even with iterative diagonalization methods, the consequence is the 
computational
time scaling bottleneck discussed already. 

Another difference is that in 
KS calculations orbital-angular-momentum-dependent (usually called non-local) 
pseudopotentials commonly are used to great benefit. In OF-DFT one
obviously cannot resort to such pseudopotentials;  
a local pseudopotential is required. Relevant consequences are 
discussed in Sect.\ V.  

Whether by use of (\ref{E6}) or (\ref{E10}), once the variational minimum
is obtained and $n$ is known for a specific ionic configuration, 
the  electronic forces on the ions  
are calculated. In KS-AIMD this done via the Hellmann-Feynman theorem 
\cite{Feynman.1939}.  In OF-DFT, the calculation is done directly from 
\cite{Perspectives}
\bea
{\mathbf F}_I&=&-\nabla_{{\mathbf R}_I}\Big({\mathcal F}[n]+ %
\int d{\mathbf r}\, v_{\mathrm{ext}}(\{{\mathbf R}\};{\mathbf r})n({\mathbf r}) %
+ E_{\mathrm{ion}}(\{{\mathbf R}\}]\Big)
\nonumber\\
&=&
-\int d{\mathbf r} n({\mathbf r}) \nabla_{{\mathbf R}_I} %
 v_{\mathrm{ext}}(\{\mathbf R\} ;{\mathbf r})  - \nabla_{{\mathbf R}_I} %
E_{\mathrm{ion}}(\{\mathbf R\}) \nonumber \\
&-& \int d{\mathbf r} \BIG{[}{14pt}\frac{\delta {\mathcal F}_s[n]}{\delta n({\bf r})}
+v_{\rm KS}([n];{\bf r})
\BIG{]}{14pt} \nabla_{{\mathbf R}_I}n({\bf r}) \; .
\label{M12a}
\eea
At electronic equilibrium for a given ionic configuration, Eq.\ (\ref{E10})
is satisfied, so at constant volume the last integral becomes 
\be
\mu \int d{\mathbf r} \nabla_{{\mathbf R}_I} n({\mathbf r}) %
= \mu \nabla_{{\mathbf R}_I} \int d{\mathbf r} n({\mathbf r}) %
= \mu \nabla_{{\mathbf R}_I} N_e = 0
\ee
which leaves 
\be
{\mathbf F}_I =
-\int d{\mathbf r}  n({\mathbf r}) %
\nabla_{{\mathbf R}_I} v_{\mathrm{ext}}(\{\mathbf R\} ;{\mathbf r})%
 - \nabla_{{\mathbf R}_I} E_{\mathrm{ion}}(\{\mathbf R\}) \; . 
\label{M12}
\ee
The corresponding contributions to the stress tensor are 
\be
\Sigma_{\alpha\beta}=\frac{1}{V} \sum_{\nu} %
\frac{\partial \Omega[n]}{\partial h_{\alpha\nu}}h_{\beta\nu}\,,
\label{StressTensor}
\ee
where $\alpha$, $\beta$ and $\nu$ are coordinate indices, $h$ is a matrix
constructed from the cell vectors, and $V$ is the cell volume 
(see \cite{Ho..Carter08}).  Immediately the pressure follows as 
\be
P = -\frac{1}{3} \mathrm{Tr} \Sigma  \; .
\label{pressure}
\ee

\section{Non-interacting free energy functionals}

Here we describe our modifications of {\sc Profess} to implement 
finite-T functionals, including GGA 
non-interacting and XC free-energy functionals, their 
functional derivatives, and stress-tensor contributions.

\subsection{Finite temperature Thomas-Fermi}

The LDA non-interacting free-energy is the finite-T TF
functional \cite{Feynman..Teller.1949}
\be
{\mathcal F}_{\mathrm{s}}^{\mathrm{TF}}[n]=\int d\mathbf{r} %
f_{\mathrm{s}}^{\mathrm{TF}}(n(\mathbf{r}),{\mathrm T}) \; ,
\label{TF}
\ee
where
\bea
f_{\mathrm{s}}^{\rm TF}(n,{\rm T})&=& \tau_0^{\mathrm{TF}}(n)\kappa(t)\,. 
\label{fTF}
\eea
The zero-T TF kernel, $\tau_0^{\mathrm{TF}}(n)$, was defined at 
Eq.\ (\ref{tauTFdefn}).  The factor $\kappa(t)$, which is a 
combination of Fermi-Dirac integrals \cite{Bartel..Durand.1985}, 
is a dimensionless function of the reduced temperature
\be
t={\mathrm T}/{\mathrm T}_{\mathrm F} := \frac{2}%
{\beta [3\pi^2n({\mathbf r})]^{2/3}} \,.
\label{tred}
\ee
Details of the structure of $\kappa$ and its behavior are in
Ref.\ \onlinecite{KarasievSjostromTrickey12A}  along with
a high-precision fit to a computationally convenient analytical form 
(see Appendix A of \cite{KarasievSjostromTrickey12A}).  The associated TF
potential and stress tensor are 
\be
v^{\mathrm{TF}}_{\mathrm{s}}([n];{\mathbf r})\equiv %
\frac{\delta{\mathcal F}_{\mathrm{s}}^{\mathrm{TF}}[n]}{\delta n({\mathbf r})}
=\frac{\partial \tau_0^{\mathrm{TF}}(n)}{\partial n}\kappa(t)
+\tau_0^{\mathrm{TF}}(n)\kappa^\prime(t)\frac{\partial t}{\partial n} \,,
\label{vTF}
\ee
and
\be
\Sigma_{\alpha\beta}^{\mathrm{TF}}=\frac{\delta_{\alpha\beta}}{V}\int %
d{\mathbf r} \Big[f_{\mathrm{s}}^{\mathrm{TF}}(n({\mathbf r}),{\mathrm T})
-n({\mathbf r})v^{\mathrm{TF}}_{\mathrm{s}}([n];{\mathbf r}) \Big] \,.
\label{stressTF_a}
\ee
Primes indicate derivatives with respect to the corresponding arguments.

\subsection{ Finite-T SGA and GGA functionals}

Well-behaved, non-interacting GGA free energy functionals 
have distinct KE and entropic contributions of the form 
\cite{KarasievSjostromTrickey12A}
\bea
{\mathcal F}_{\mathrm{s}}^{\mathrm{GGA}}[n,{\mathrm T}]&=& \int d{\mathbf r} %
\tau_0^{\mathrm{TF}}(n)\xi(t) F_{\tau}(s_{\tau}) \nonumber\\
&-& \int d{\mathbf r}
\tau_0^{\mathrm{TF}}(n)\zeta(t) F_{\sigma}(s_{\sigma}) \; .
\label{FGGA}
\eea
Here
\bea
\xi(t)&=&\kappa(t)-t\frac{\partial \kappa(t)}{\partial t}
\nonumber\\
\zeta(t)&=&-t\frac{\partial \kappa(t)}{\partial t}\; .
\label{xizeta}
\eea
$\kappa(t)$ is as before. 
$F_{\tau}$ and $F_{\sigma}$ are
the non-interacting KE and entropic 
enhancement factors. They depend upon two distinct 
T-dependent reduced density gradients, namely 
\bea
s_{\sigma}(n,\nabla n,t)&=&s(n,\nabla n)%
\Big(\frac{t{\rm d} \tilde h(t)/{\rm d}t}{\zeta(t)}\Big)^{1/2}\,, \nonumber\\
s_{\tau}(n,\nabla n,t)&=&s(n,\nabla n)%
\Big(\frac{\tilde h(t)-t{\rm d}\tilde h(t)/{\rm d}t}{\xi(t)}\Big)^{1/2}\, 
\label{s_s}
\eea
where
\be
s(n,\nabla n)=
\frac{1}{2(3\pi^2)^{1/3}} \, \frac{|\nabla n|}{n^{4/3}} \; 
\label{s_0defn}
\ee
is the reduced density gradient familiar from T = 0 K GGA XC functionals.  
The function $\tilde h$ in Eq.\ (\ref{s_s}) is another
combination of Fermi-Dirac integrals for which an analytical fit 
is provided in Appendix A of Ref.\ \onlinecite{KarasievSjostromTrickey12A}.

Functionals of the form of Eq.\ (\ref{FGGA}) which we have added in
{\sc Profess} include: \\
(i) the purely non-empirical functional obtained via a new constraint-based 
parametrization scheme \cite{VT84F}
\bea
F_{\tau}^{\mathrm{VT84F}}(s_{\tau})&=&1-\frac{\mu s_{\tau}^2e^{-\alpha s_{\tau}^2}}{1+\mu s_{\tau}^2} 
\nonumber\\
&+& (1-e^{-\alpha s_{\tau}^{m/2}})
(s_{\tau}^{-n/2}-1)+\frac{5}{3}s_{\tau}^2
\nonumber\\
F_{\sigma}^{\mathrm{VT84F}}(s_{\sigma})&=&2-F_{\tau}^{\mathrm{VT84F}}(s_{\sigma})
\,,
\label{VT84F}
\eea
with $m=8$, $n=4$, $\mu=2.778$ and $\alpha=1.2965$;\\
(ii) the mildly empirical (from a small set of molecular data) 
two-parameter (KST2) \cite{KarasievSjostromTrickey12A} functional 
\bea
F_{\tau}^{\mathrm{KST2}}(s_\tau)&=&1+\frac{C_1s_\tau^2}{1+a_1s_\tau^2}
\nonumber\\
F_{\sigma}^{\mathrm{KST2}}(s_\sigma)&=&2-F_{\tau}^{\mathrm{KST2}}(s_\sigma)
\; ;
\label{KST2F}
\eea
with constants $C_1=2.03087$, $a_1=0.29424$;\\
(iii) the finite-T extension of the zero-T APBEK 
functional \cite{CFDS11} given 
by Eq.\ (\ref{KST2F}) with $C_1=0.23889$ and $a_1=C_1/0.804$  
(see Ref.\ \cite{VT84F}); \\
(iv) the finite-T extension of the Tran-Wesolowski \cite{TranWesolowski02} (TW)
ground state functional,   
given by Eq.\ (\ref{KST2F}) with $C_1=0.2319$ and $a_1=0.2748$ 
(again see Ref.\ \cite{KarasievSjostromTrickey12A}); \\
(v) the finite-T SGA, also known as the gradient-corrected TF
model \cite{Perrot.1979},
\bea
F_{\tau}^{\mathrm{SGA}}(s_\tau)&=&1+\mu^{\mathrm{SGA}} s_\tau^2
\nonumber\\
F_{\sigma}^{\mathrm{SGA}}(s_\sigma)&=&2-F_{\tau}^{\mathrm{SGA}}(s_\sigma)
\; ,
\label{SGA-F}
\eea
with $\mu^{\mathrm{SGA}}=5/27$;
\\
(vi) the empirical combination of the von Weizs\"acker\cite{Weizsacker} 
functional, Eq.\ (\ref{T_Wdefn}), and the finite-T Thomas-Fermi 
(ftVWTF) functional \cite{KarasievSjostromTrickey12A}
given again by Eq.\ (\ref{SGA-F}) but with $\mu^{\mathrm{VWTF}}=5/3$ instead
of $\mu^{\mathrm{SGA}}$. 

The potential in the Euler equation that arises from any of the 
${\mathcal F}_{\mathrm{s}}^{\rm GGA}$ functionals 
can be evaluated using the generic equation for the functional derivative
of a functional dependent on $n$ and $\nabla n$:
\begin{widetext}
\bea
v^{\mathrm{GGA}}_{\mathrm{s}}([n];{\mathbf r})&\equiv& %
\frac{\delta{\mathcal F}_{\mathrm{s}}^{\mathrm{GGA}}[n]}{\delta n({\mathbf r})} %
\nonumber\\
&=&\frac{\partial \tau_0^{\rm TF}(n)}{\partial n}\xi(t)F_{\tau}(s_{\tau})
+ \tau_0^{\rm TF}(n)\xi'(t)\frac{\partial t}{\partial n}F_{\tau}(s_{\tau})
+ \tau_0^{\rm TF}(n)\xi(t)\frac{\partial F_{\tau}(s_{\tau})}{\partial s_{\tau}^2}
\Big(\frac{\partial s_{\tau}^2}{\partial n}+\frac{\partial s_{\tau}^2}{\partial t}\frac{\partial t}{\partial n}\Big)
\nonumber\\
&-&\frac{\partial \tau_0^{\rm TF}(n)}{\partial n}\zeta(t)F_{\sigma}(s_{\sigma})
+ \tau_0^{\rm TF}(n)\zeta'(t)\frac{\partial t}{\partial n}F_{\sigma}(s_{\sigma})
+ \tau_0^{\rm TF}(n)\zeta(t)\frac{\partial F_{\sigma}(s_{\sigma})}{\partial s_{\sigma}^2}
\Big(\frac{\partial s_{\sigma}^2}{\partial n}+\frac{\partial s_{\sigma}^2}{\partial t}\frac{\partial t}{\partial n}\Big)
\nonumber\\
&-&\nabla \cdot \Big(
\tau_0^{\rm TF}(n)\xi(t)\frac{\partial F_{\tau}(s_{\tau})}{\partial s_{\tau}^2}\frac{\partial s_{\tau}^2}{\partial \nabla n} 
- \tau_0^{\rm TF}(n)\zeta(t)\frac{\partial F_{\sigma}(s_{\sigma})}{\partial s_{\sigma}^2}\frac{\partial s_{\sigma}^2}{\partial \nabla n}
\Big)  \,.
\label{vftGGA}
\eea

Rather than writing a single complicated expression for 
$v^{\mathrm{GGA}}_{\mathrm s}$ and evaluating it in {\sc Profess}, we take 
advantage of the structural commonality of
all GGAs for the non-interacting free energy.  The RHS of Eq.\ (\ref{vftGGA})
shows that there are only four factors which depend on a specific GGA,
$F_\tau$, $F_\sigma$, 
$\partial F_\tau/\partial s_\tau^2$, and $\partial F_\sigma/\partial s_\sigma^2$.
All the other contributions are generic for GGAs, 
e.g. $\zeta$, $\partial t/\partial n$, $\partial s_\tau^2/\partial n$, etc.  
The code is constructed to evaluate both the four specific contributions and 
all the generic ones individually, then assemble the result.  
The last line of (\ref{vftGGA}) 
is calculated in reciprocal space, then inverse Fourier transformed.  

The ftGGA stress tensor components are
\bea
\Sigma_{\alpha\beta}^{\rm GGA}&=&\frac{\delta_{\alpha\beta}}{V}\int d{\mathbf r} 
\Big[f_s^{\rm GGA}([n];{\bf r})-n({\bf r})v^{\rm GGA}_{\mathrm{s}}([n];{\bf r}) \Big]
\nonumber\\
&-& \frac{1}{V} \int d{\mathbf r} 
\Big[ \Big( 
\tau_0^{\rm TF}(n)\xi(t)\frac{\partial F_{\tau}(s_{\tau})}{\partial s_{\tau}^2}\frac{\partial s_{\tau}^2}{\partial |\nabla n|} 
- \tau_0^{\rm TF}(n)\zeta(t)\frac{\partial F_{\sigma}(s_{\sigma})}{\partial s_{\sigma}^2}\frac{\partial s_{\sigma}^2}{\partial |\nabla n|}
\Big)
\frac{\nabla_{\alpha}n({\bf r}) \nabla_{\beta}n({\bf r})}{|\nabla n|}
\Big] \,,
\label{stressTF_b}
\eea
\end{widetext}
where
\be
 f_s^{\mathrm{GGA}}([n];{\bf r})=\tau_0^{\rm TF}(n)\xi(t) F_{\tau}(s_{\tau})- %
\tau_0^{\mathrm{TF}}(n)\zeta(t) F_{\sigma}(s_{\sigma})
\label{fsidentity}
\ee
and $\alpha$ and $\beta$ are Cartesian coordinate indices.

\section{Exchange-correlation free energy functionals}

Both the {\sc Profess} and {\sc Quantum Espresso} packages have 
standard ground state LDA \cite{PZ81} and GGA \cite{PBE96} XC functionals 
implemented.  For $T >$ 0 K, the explicit
T-dependence of the XC free-energy may be important.  See
discussions in Refs.\ \onlinecite{KarasievSjostromTrickey2012PRE,BrownEtAl2013}.
Our enhancements of {\sc Profess} and {\sc Quantum Espresso},  
include three explicitly T-dependent XC functionals, though with
the recent advent of the first one \cite{LSDA-PIMC}, the other two may be mostly
of value for checking against earlier literature.

\subsection{LDA parametrized from path-integral Monte Carlo data}

Recently we and a co-author presented a finite-T local 
spin-density approximation (LSDA) XC free-energy functional
\cite{LSDA-PIMC} obtained via accurate parametrization 
of  {\it first principles} restricted 
path-integral Monte Carlo simulation data for the 3D homogeneous electron
gas at finite T. The XC free-energy per particle is given
by a function of $(r_{\mathrm{s}},t)$, with $r_{\mathrm s} =(3/4\pi n)^{1/3}$, 
and $t$ defined by Eq. (\ref{tred}) for the spin-unpolarized case, and  
$t={\mathrm T}/{\mathrm T}_{\mathrm F}^{\mathrm{p}}\equiv\frac{2}%
{\beta [6\pi^2n({\mathbf r})]^{2/3}}$ for the fully spin-polarized case,
\be
f_{\mathrm{xc}}^{\mathrm{u/p}}(r_{\mathrm{s}},t)=-\frac{1}{r_{\mathrm{s}}}
\frac{\omega_{\mathrm{u/p}} a(t)+b_\mathrm{u/p}(t)r_{\mathrm{s}}^{1/2}+c_\mathrm{u/p}(t)r_{\mathrm{s}}}
{1+d_\mathrm{u/p}(t)r_{\mathrm{s}}^{1/2}+e_\mathrm{u/p}(t)r_{\mathrm{s}}}
\,,
\label{fit2}
\ee
where $\omega_{\mathrm{u}}=1$ and $\omega_{\mathrm{p}}=2^{1/3}$
for the spin-unpolarized and fully-polarized cases respectively.
The functions $a(t)$ and $b_\mathrm{u/p}(t) - e_\mathrm{u/p}(t)$ 
%
are given in \cite{LSDA-PIMC}.
In the small-$r_{\mathrm{s}}$ limit, Eq.\ (\ref{fit2}) reduces to the
LSDA finite-T exchange defined by exact scaling relations for X and fitted in 
Ref.\ \cite{PDW84}, $f_{\mathrm{x}}^{\mathrm{u/p}}(r_{\mathrm{s}},t)=-\omega_{\mathrm{u/p}} a(t)/r_{\mathrm{s}}$.
The corresponding XC functional derivative is
\bea
v_{\mathrm{xc}}^{\mathrm{u/p}}([n];{\mathbf{r}})&\equiv&
\frac{\delta \Big(\int d{\mathbf{r}} n f_{\mathrm{xc}}^{\mathrm{u/p}}(r_{\mathrm{s}},t)\Big)}
{\delta n({\mathbf{r}})}=f_{\mathrm{xc}}^{\mathrm{u/p}}(r_{\mathrm{s}},t)
\nonumber\\
&+&
n\Big[\frac{\partial f_{\mathrm{xc}}^{\mathrm{u/p}}(r_{\mathrm{s}},t)}{\partial r_{\mathrm{s}}}
\frac{\partial r_{\mathrm{s}}}{\partial n}
+
\frac{\partial f_{\mathrm{xc}}^{\mathrm{u/p}}(r_{\mathrm{s}},t)}{\partial t}
\frac{\partial t}{\partial n}\Big]\,,
\label{vxc1}
\eea
where $\partial r_{\mathrm{s}}/\partial n=-r_{\mathrm{s}}/3n$ and 
$\partial t/\partial n=-2t/3n$ independent of spin-polarization. 

The stress tensor for the XC free-energy is given by the 
standard expression for an LDA XC functional
\be
\Sigma_{\mathrm{xc},\alpha\beta}^{\mathrm{u/p}}=\frac{\delta_{\alpha\beta}}{V}\int %
d{\mathbf r} \Big[n({\mathbf r})f_{\mathrm{xc}}^{\mathrm{u/p}}(r_{\mathrm{s}},t)
-n({\mathbf r})v^{\mathrm{u/p}}_{\mathrm{xc}}([n];{\mathbf r}) \Big] \,.
\label{stressLSDAxc}
\ee

Only the spin-unpolarized version 
of the functional
given by Eq.\ (\ref{fit2})
is implemented in our modifications of the current version of 
{\sc Profess} and {\sc Quantum Espresso}.

\subsection{RPA functional}

The functional developed by Perrot and Dharma-wardana (PDW84) \cite{PDW84}
for the fully unpolarized case
combines finite-T parametrized LDA exchange 
(in the form of the small $r_{\mathrm s}$ limit of Eq.\ (\ref{fit2}))  
with correlation treated 
via the random-phase approximation (RPA).  
The X free energy per electron in that functional is defined as 
\be
f_{\mathrm x}^{\mathrm{PDW84}}(r_{\mathrm{s}},t)\equiv f_{\mathrm x}^{\mathrm{u}}(r_{\mathrm{s}},t)=-\frac{1}{r_{\mathrm{s}}}a(t)
\,. 
\label{fx}
\ee
%
%
Perrot and Dharma-wardana also used the form of Eq.\ (\ref{fx})
to fit the corresponding exchange potential $v_{\mathrm x}$ independently. 
Elsewhere \cite{KST-PDW84} we have shown that for a 
consistent pressure calculation, such an independent fit of $v_{\mathrm x}$ 
should not be used.  Rather, $v_{\mathrm x}$ should be calculated 
via direct evaluation of the functional derivative of the
specified X functional via use of an equation that corresponds to 
Eq.\ (\ref{vxc1}).
The PDW84 correlation contribution is 
\bea
f_{\mathrm c}^{\mathrm{PDW84}}(r_{\mathrm{s}},t)&=&
\epsilon_{\mathrm c}^{\mathrm{LDA}}(r_{\mathrm{s}})(1+c_1t+c_2t^{1/4})\exp(-c_3t)
\nonumber\\
&-&0.425437~(t/r_{\mathrm{s}})^{1/2}
\nonumber\\
&\times& \tanh(1/t)\exp(-c_4/t)
\,,
\label{fc}
\eea
where the $c_i$ are explicit functions of 
$r_{\mathrm{s}}$ and 
$\epsilon_{\mathrm c}^{\mathrm{LDA}}(n)$ is the zero-T LDA correlation energy 
per electron given by the Vosko, Wilk, and Nusair (VWN) parametrization 
\cite{VWN80} or by the re-parametrized Hedin-Lunqvist (rHL)
local form \cite{HL.1971,McDDWG.1980}
(see Eqs.\ (3.7), (3.9)-(3.10) of Ref.\ \onlinecite{PDW84}).
The correlation potential is the functional derivative given again by
the analogue of Eq.\ (\ref{vxc1}).
%
The stress tensor for the PDW84 XC functional is given by the  
spin-unpolarized version of Eq.\ (\ref{stressLSDAxc}).

\subsection{Classical map functional}

Ref.\ \onlinecite{PDW2000}  presented an XC free-energy functional
(hereafter denoted PDW00) built by mapping between 
the quantum system of interest 
and a more tractable classical system.
We have implemented that functional via the fits 
provided in Ref.\ \onlinecite{PDW2000} for the XC 
free energy per electron. For completeness that fit is
\begin{align}
  f^{\rm PDW00}_{\rm xc} (r_{\mathrm{s}}, \mathrm{T}) &= (\varepsilon_{\rm xc}(r_{\mathrm{s}},0)-P_1)/P_2, 
   \label{eq:pdw2000} \\
  P_1&=(A_2u_1+A_3u_2)\mathrm{T}^2 +A_2u_2\mathrm{T}^{5/2}, \\
  P_2&=1+A_1\mathrm{T}^2+A_3\mathrm{T}^{5/2}+A_2\mathrm{T}^3, 
\end{align}
%
where $A_{1}$, $A_{2}$, $A_{3}$, $u_{1}$, and  $u_{2}$ are functions of 
$r_{\mathrm{s}}$ (see definitions in Ref.\ \onlinecite{PDW2000}). 
In addition, 
$\varepsilon_{\rm xc}(r_{\mathrm{s}},0)$ is the zero-T XC LDA energy per particle 
given by the PZ parametrization of QMC results \cite{PZ81}. 
Also we have implemented the 
XC potential which follows from  exact differentiation of 
Eq.\ \ref{eq:pdw2000} as
%
\bea
v_{\mathrm xc}^{\rm PDW00}(n({\mathbf r}),\mathrm{T})&\equiv&
\frac{\delta 
\Big(\int d{\mathbf r} n\,f_{\mathrm xc}^{\rm PDW00}(n,\mathrm{T}) \Big)
} {\delta n({\mathbf r})}  \nonumber\\
= f_{\mathrm xc}^{\rm PDW00}(n,\mathrm{T}) 
&+& n\, \frac{\partial f_{\mathrm xc}^{\rm PDW00}(n,\mathrm{T})}{\partial n}
\,.
\label{vxc}
\eea
The stress tensor for the PDW00 functional again is given by the 
spin-unpolarized version of Eq.\ (\ref{stressLSDAxc}).

%

\section{Implementational Aspects}
\vspace*{-1pt}
\subsection{Local pseudopotentials}

Efficient use of Fourier-based numerical methods in OF-DFT computation
as well as alleviation of difficulties introduced by the bare Coulomb 
nuclear-electron interaction make it 
desirable to regularize the singular external potential $v_{\mathrm{ext}}$.  
In the KS context, regularization often 
is accomplished with non-local pseudopotentials. The
essential feature of such potentials, for this discussion, is their  
explicit dependence upon atomic orbital angular momentum, which makes 
them intrinsically incompatible with OF-DFT.  Local
pseudopotentials (LPPs) are required.  The LPPs in a reciprocal 
space representation used by {\sc Profess} are described in 
Refs.\ \onlinecite{Ho..Carter08,Hung..Carter10}.

More recently,  we have developed LPPs in 
both direct and reciprocal space for Li as described in Ref.\ 
\onlinecite{KarasievTrickeyCPC2012}.  A hydrogen LPP intended for 
finite-T WDM applications was 
developed and tested in Ref.\ \onlinecite{KarasievSjostromTrickey12A}.
It has the reciprocal space form of the Heine-Abarenkov model 
\cite{Heine.Abarenkov.1964,Goodwin..Heine.1990}, namely
\bea
v_{\mathrm{HA}}(q)&=&\frac{-4\pi}{V q^2}[(Z-Ar_{\mathrm c}){\,\cos}%
(qr_{\mathrm c})  \nonumber \\
&& +(A/q){\,\sin}(qr_{\mathrm c})] f_{\mathrm HA}(q) \; , \nonumber \\
f_{\mathrm HA}(q) &:=& \exp{(-(q/q_{\mathrm c})^6)}  \;.
\label{E4A}
\eea
The parameters for H are $r_{\mathrm c}=0.25$ bohr, $A=6.18$ hartree and  
$q_{\mathrm c}=29.97$ bohr$^{-1}$. 
The parameters for the Al atom,
developed in Ref.  \onlinecite{Goodwin..Heine.1990}, are 
$r_{\mathrm c}=1.15$ bohr, $A=0.1107$ hartree and 
$q_{\mathrm c}=3.5$ bohr$^{-1}$.
All of these LPPs are available for 
download \cite{web-QTP-PP}.  The reciprocal space form is incorporated
in {\sc Profess}$@${\sc Q-Espresso}.

\subsection{Modifications of {\sc Profess}}


All of the functionals described in the foregoing Sections are implemented 
in our modification of 
the {\sc Profess} code, version 2.0. A short description of the most important subroutines and functions added
to the modified version is given in Table \ref{table:modif-profess}.
 A new keyword, \texttt{TEMP \{real\}}, defines \texttt{TEMP}erature 
in Kelvin.
Also we found it useful, especially for warm dense matter simulations, to define
a new input parameter, \texttt{cellScale \{real\}} at the end of the first 
section of the geometry (\texttt{.ion}) file, as follows:
\begin{verbatim}
%BLOCK LATTICE_CART
  ... 
  ...
  ...
  cellScale
%END BLOCK LATTICE_CART
\end{verbatim}
This parameter defines a scaling factor for all lattice vectors and for all 
atomic coordinates. Its default value is 1.0.

\subsection{New XC functionals implemented in {\sc Quantum Espresso}}

We have implemented the explicitly T-dependent XC functionals described 
in the preceding Section in our modification of {\sc Quantum Espresso} 
vers.\ 5.0.3. 
Table \ref{table:modif-QE} lists the added subroutines and keywords. 
These functionals can be used in standard KS calculations (static or AIMD) 
with {\sc Quantum Espresso}.
Note that the exchange-correlation internal and entropic contributions 
for the KSDT functional based on the PIMC simulation data \cite{LSDA-PIMC},  
(\texttt{input\_dft='KSDT'} keyword) are calculated separately and 
the energy contributions listed in the standard output are changed accordingly
(see next Subsection for details).


\begin{table*}
  \caption{Description of new functions and subroutines implemented in %
the modified version of {\sc Profess} vers.\ 2.0.}
  \begin{ruledtabular}
    \begin{tabular}{lll}
Keyword & Calculates  & function/subroutine \\
        \hline
\texttt{KINE MCPBE2}  & PBE2  zero-T KE \cite{Perspectives,PRB80} and functional derivative\cite{KarasievTrickeyCPC2012}.  & \verb|mcPBE2PotentialPlus| \\
 --- ---              & The associated stress tensor $\Sigma_{\alpha\beta}^{\mathrm{PBE2}}$.  & \verb|mcPBE2Stress|\\
\hline
\texttt{KINE PBETW}  & TW zero-T KE \cite{TranWesolowski02} and functional derivative\cite{KarasievTrickeyCPC2012}.  & \verb|PBETWPotentialPlus| \\
 --- ---            & The associated stress tensor  $\Sigma_{\alpha\beta}^{\mathrm{TW}}$.  & \verb|PBETWStress|\\
\hline
      --- ---         & The function $f(y)$\tablenotemark[1] and derivatives, function $h\equiv\tilde h/72$ and derivatives \cite{Perrot.1979}. & \verb|FPERROT| and \verb|FPERROT2|\\
\hline
\texttt{KINE TTF} & TF free energy, entropic contribution, and functional derivative.  & \verb|TTF1PotentialPlus| \\
    --- ---        & The associated stress tensor $\Sigma_{\alpha\beta}^{\mathrm{TF}}$.  & \verb|TTF1Stress| \\
\hline
\texttt{KINE VT84F} & VT84F \cite{VT84F} free energy, entropic contribution, and functional derivative. & \verb|TVTPotentialPlus| \\
                       & VT84F stress tensor. & \verb|TVTStress| \\
\hline
\texttt{KINE HVWTF} \& & SGA and VWTF free energy, entropic contribution, and functional & \verb|TPBE2PotentialPlus| \\
\texttt{PARA MU 'value'} & derivative \cite{KarasievSjostromTrickey12A}, with the second term in Eq. (\ref{SGA-F}) & \\
\vspace{4pt}
& multiplied by value of parameter $\mu$\tablenotemark[2].  & \\ 
\texttt{KINE KST2} & KST2 \cite{KarasievSjostromTrickey12A} free energy, entropic contribution, and functional derivative. & \\
\texttt{KINE PBETWF} & TW free energy, entropic contribution, and functional derivative. & \\
\vspace{4pt}
\texttt{KINE APBEF} & APBEF \cite{VT84F} free energy, entropic contribution, and functional derivative. & \\
& SGA, VWTF, KST2,TW, and APBEF stress tensor. & \verb|TPBE2Stress| \\
\hline
\texttt{EXCH KSDT} & LDA XC free-energy based on PIMC simulation data \cite{LSDA-PIMC}. & \verb|KSDTXCPotentialPlus| \\
   --- ---    & LDA XC internal energy based on PIMC simulation data. & \verb|KSDT_EXC|  \\
 --- ---            & Stress tensor corresponding to LDA XC free-energy.  & \verb|KSDTXCStress|  \\
\hline
\texttt{EXCH PDWX+NONE} & PDW84 exchange free energy.  & \verb|PDWXPotentialPlus| \\
\texttt{EXCH ...+PD84} & PDW84 correlation free energy \tablenotemark[3]. & \verb|PD84CPotentialPlus|\\
\texttt{EXCH PDW00XC} & PDW00 exchange-correlation free energy. & \verb|PD00XCPotentialPlus|\\
--- --- & PDW84 exchange contribution to the stress tensor. & \verb|PDWXStress| \\ 
--- --- & PDW84 correlation contribution to the stress tensor. & \verb|PDWCStress| \\
--- --- & PDW00 stress tensor. & \verb|PD00XCStress| \\
    \end{tabular}
\tablenotetext[1]{$\kappa(t)=(5/3)tf(y(t))$, see also Refs. \cite{KarasievSjostromTrickey12A} and \cite{Perrot.1979}.}
\tablenotetext[2]{SGA free energy Eq. (\ref{SGA-F}) corresponds to $\mu=\mu^{\mathrm{SGA}}=5/27$, and VWTF corresponds to 
$\mu=\mu^{\mathrm{VWTF}}=5/3$.}
\tablenotetext[3]{Use \texttt{EXCH PDWX+PD84} for PDW84 exchange and correlation.}
  \end{ruledtabular}
  \label{table:modif-profess}
\end{table*}

\begin{table*}
  \caption{Description of new functions and subroutines implemented in %
{\sc Quantum Espresso} vers.\ 5.0.3.}
  \begin{ruledtabular}
    \begin{tabular}{lll}
Keyword & Calculates & function/subroutine \\
\hline
\texttt{input\_dft='KSDT'} & LDA XC free-energy based on PIMC simulation data. & \verb|fxc_ksdt_0|\\
                            & LDA XC internal energy based on PIMC simulation data. & \verb|exc_ksdt_0|\\
\hline
 \texttt{input\_dft='TXC0'} & PDW00 XC free-energy. & \verb|pdw00xc|\\
\hline
\texttt{input\_dft='PDWX+...'} & PDW84 exchange free energy. & \verb|pdwx_t|\\
\texttt{input\_dft='...+PD84'} & PDW84 correlation free energy. & \verb|pdw84|\\
\texttt{input\_dft='...+PDW0'} & PDW00 correlation free energy. & \verb|pdw00|\\
    \end{tabular}
  \end{ruledtabular}
  \label{table:modif-QE}
\end{table*}

\subsection{ {\sc Profess}$@${\sc Q-Espresso} interface}

The implementation of {\sc Profess}$@${\sc Quantum Espresso} uses 
the modified {\sc Profess} compiled as an external library for 
{\sc Quantum Espresso}.
{\sc Profess@Q-Espresso} requires standard input files for
both the {\sc Quantum Espresso} \texttt{PWscf} package  
and for {\sc Profess}.  In addition, 
for each step (ionic configuration) of an OF-DFT AIMD calculation,   
{\sc Profess} is called from the {\sc Quantum Espresso}  
\texttt{PWscf} main program via the interface instead of executing
the normal KS procedure within \texttt{PWscf}.  The first such
call is special: 
\begin{verbatim}
  ...
  IF(useofdft) THEN
    if(istep.eq.0) CALL create_ofdftgeom()
    CALL ofdft_driver()
  ENDIF
  ... 
\end{verbatim} 
Here \verb|useofdft {logical}|  is a new keyword added 
in  \verb|Namelist: CONTROL| with default value 
\verb|useofdft=.false.| 
The second \verb| if | statement takes care of initialization 
of {\sc Profess} upon its first use.   The interface creates
the {\sc Profess} geometry input file, \texttt{inp.ion}, on  
the basis of input provided for
{\sc Quantum Espresso}.  That file includes the  
description of the cell vectors, atom types, and intra-cell atom 
coordinates (see the subroutine \texttt{create\_ofdftgeom}). 
Required initializations of {\sc Profess} with data provided in the input file
\texttt{ofdft.inpt} also are
done at that first call.  

At all subsequent MD steps,    
data exchange between {\sc Quantum Espresso} and {\sc Profess} 
via the interface is minimal.  After each MD step, the 
current ionic (nuclear) coordinates $\{{\mathbf R}\}$ 
(\texttt{tau\_ofdft}) and the number of atoms (\texttt{nat}) 
are passed to {\sc Profess} from {\sc Quantum Espresso} via 
the interface.  Prior to the next OF-DFT AIMD step, 
the interface transmits the following items 
from {\sc Profess} to {\sc Quantum Espresso}:
the current contributions to the total free energy (\texttt{etot}), 
electronic internal energy (\texttt{einternel}), stress-tensor components
(in principle; at present it returns only the pressure \texttt{pressure})
and new Born-Oppenheimer forces $\{{\bf F}\}$ (\texttt{force\_ofdft}) for
the next step.  All of this is achieved from \texttt{ofdft\_driver} via 
\begin{verbatim}
  ...
  CALL profess(nat,tau_ofdft,force_ofdft, &
               pressure,etot,einternel)
  ... 
\end{verbatim}
The flow of control was 
shown in Fig.\ \ref{P-QE}. 
The ionic coordinates measured as fractions of cell constants (and
denoted as fractional atomic coordinates) at each MD step are stored 
in the \texttt{md.xyz} file for subsequent visualization.  Note that
{\sc Quantum Espresso} uses Rydberg atomic units, while
{\sc Profess} uses Hartree atomic units internally and reports
results in eV and {\AA}ngstroms.  The interface takes account of
this unit system difference.  

The modified version of {\sc Profess} also may be compiled 
as a stand-alone package from the same source files.  Doing so enables 
OF-DFT calculations for either single-point or static geometry optimization 
without
need of the interface to {\sc Quantum Espresso}.  The new subroutines in
the modified version have the same parallelization, implemented through
domain decomposition using MPI, though it is a development version
with some inefficiently implemented parts.  

Because the implementation includes XC functionals with explicit temperature
dependence, both the internal and entropic contributions  of the XC free-energy
are calculated and the output is changed
correspondingly. Here is a sample of modified 
{\sc Quantum Espresso} output during OF-DFT MD simulation with the 
KSDT T-dependent LSDA XC functional
\begin{verbatim}
  ...
 OFDFT: (kbar) P=   7192.78778379771     
 OFDFT: (Ry) Fxc=  -96.3275434867815     
 OFDFT: (Ry) Exc=  -106.396016259169
  ...
 kinetic energy (Ekin) =       156.27791852 Ry
 temperature           =    129523.89686036 K 
 Ekin + Etot (const)   =       -28.87878502 Ry
 free energy Etot      =      -185.15670354 Ry
 EinternEl(PR)             =   -30.86970711 Ry
 Eint=Ekin + EinternEl(PR) =   125.40821141 Ry
 smearing contrib.(-TS)(PR)=  -154.28699643 Ry
  ... 
\end{verbatim}
Shown are values of the pressure  \texttt{P} from electronic structure
calculations (ideal gas ionic contribution is not included), 
XC free energy \texttt{Fxc},
XC internal energy \texttt{Exc}, ionic kinetic energy  \texttt{Ekin},
ionic temperature, sum of ionic-kinetic and free-energy \texttt{Ekin + Etot},
free-energy \texttt{Etot}, internal energy  \texttt{EinternEl},
sum of ionic-kinetic and internal energy \texttt{Eint=Ekin + EinternEl}
and the entropic contribution corresponding to the 
non-interacting term $-\mathrm{T}\mathcal{S}_{\mathrm{s}}$
and XC contribution $\mathcal F_{\mathrm{xc}}-\mathcal E_{\mathrm{xc}}$.
The output for Kohn-Sham MD with T-dependent XC has the same quantities 
in slightly different  order. Calculation of internal XC 
contributions for the PDW84 and PDW00 T-dependent
functionals is not implemented at present, hence the partition 
of the free-energy (\texttt{Etot})
into the internal (\texttt{EinternEl}) and
entropic contributions (\texttt{-TS}) is not correct for these functionals; 
only the free energies \texttt{Etot} and \texttt{Ekin + Etot} have meaningful  
values.

\subsection{User implementation}

In addition to the added functions and subroutines already described,
the down-loadable material for user implementation of these 
modifications includes a basic \texttt{README} 
file.  It lists patches which are required in the two codes, 
describes precisely how to install them, 
and includes scripts to run a few 
example static, OF-DFT AIMD and KS AIMD calculations as basic tests.

An important technical oddity is that {\sc Profess} and {\sc Quantum Espresso}
require distinct, \emph{incompatible} versions of the Fourier transform
package \texttt{fftw} \cite{fftw}.  Thus, the downloads include \texttt{fftw} version 2.1.5
modified specifically for use with {\sc Profess} and to avoid multiple
definition problems caused by simultaneous use of version 3.3 by 
{\sc Quantum Espresso}.

\section{Results and discussion}

In this Section we illustrate use of these added capabilities with
two examples.  First are timing and scaling comparisons 
for static lattice aluminum in the face-centered cubic (fcc) phase. 
Second are timing and scaling comparisons 
for AIMD simulations of hydrogen with OF-DFT AIMD via 
{\sc Profess@Q-Espresso} and KS AIMD with {\sc Quantum Espresso} 
at temperatures up to ${\mathrm T}=4,000,000$ K. 
All parallelized benchmark calculations 
used the high-performance computing cluster (2.4-2.8 GHz AMD Opteron) 
at University of Florida.  Serial tests were on a single 3.2 GHz Intel $i5$ 
CPU. 

Procedural context is helpful for interpreting the results.  In
ordinary practice, we do the OF-DFT AIMD simulations on 8-32 cores. To
complete 6,000 MD steps with 128 H atoms in the simulation cell on a
$64\times64\times64$ numerical grid typically takes between a few
hours and a few days depending on simulation details such as the CPU
speed, number of cores, thermodynamic conditions, and the choice of
functional. A KS AIMD simulation with the same number of steps on 32
cores for the same number of H atoms at material density 0.983
g/cm$^3$ on a $3\times3\times3$ {\textbf k}-mesh at ${\mathrm T}=30,000$ K 
takes about one month.  For  $\Gamma$ point KS calculations at 125,000 K and
180,000 K, the corresponding timings are about one and two months
respectively.  

\subsection{Static calculations - scaling}

\begin{figure}
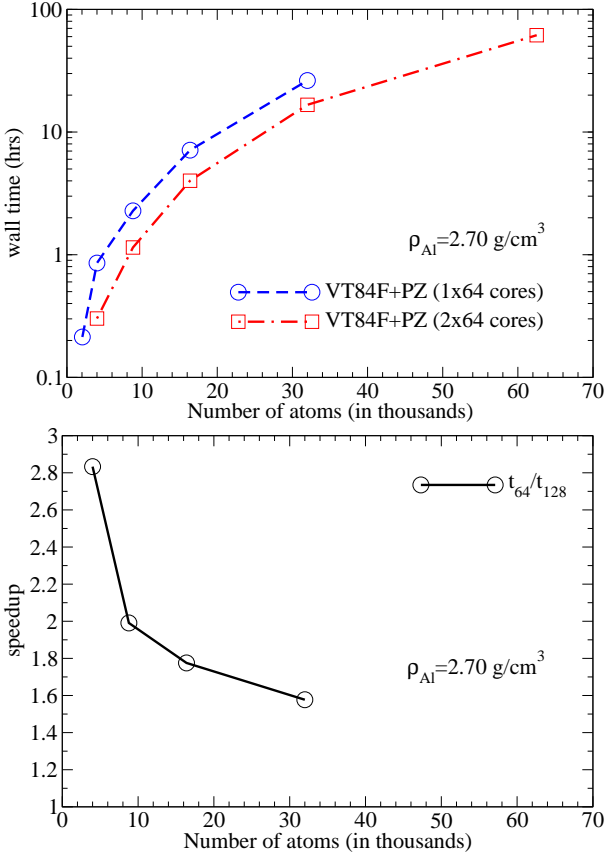

\includegraphics*[angle=-00,width=8.0cm]{cputime-1step-64cpu-fcc-Al.OFDFT.eps}
\includegraphics*[angle=-00,width=8.0cm]{parallel-efficiency-64-128.eps}
\caption{
Upper panel: execution time for OF-DFT static calculations with VT84F 
non-interacting free-energy functional and PZ XC for fcc-Al
at material density $\rho_{\rm Al}=2.70$ g/cm$^3$ and  
electronic temperature ${\mathrm T}_{\mathrm{el}}=100$ K
with 2048, 4000, 8788, 16383, 32000, and 62500 atoms in the supercell.
Calculations used  1 node $\times$ 64 cores and 2 nodes 
$\times$ 64 cores of AMD Opteron 2.4 GHz CPUs.
Lower panel: speedup from 64 to 128 cores as a function of number of atoms.
}
\label{time-fcc-Al}
\end{figure}

Static OF-DFT calculations for fcc-Al were done with between
2048 and 62500 atoms in a supercell. These used the modified 
{\sc Profess} code alone; the interface to {\sc Quantum-Espresso} and 
{\sc Quantum-Espresso} itself are un-needed.
Figure \ref{time-fcc-Al}
shows the wall-clock time as a function of number of atoms for
the VT84F non-interacting free energy functional and the PZ ground-state
XC functional.  Calculations show almost linear scaling $O(N\ln N)$ 
with system size \cite{CPL.Carter.2009}.   The computational time on 
128 cores for a single density optimization with 32,000 atoms is about 
17 hours. The lower panel of that figure shows the speedup in going from 
64 to 128 cores; at 32,000 atoms the speedup is about 80\% of optimal. 
The super-linear speedup at 4000 atoms apparently is a consequence of
being able to keep the relatively small problem primarily in cache,
something which is not possible at 64 cores.  
  
A significant sidelight is that the non-empirical APBEK 
non-interacting functional not only fails to predict physically meaningful
results (no binding at T=0 K), it sometimes exhibits bad numerical 
convergence, apparently because of improper large-$s$ behavior \cite{VT84F}.
Our mildly empirical ground-state non-interacting functional PBE2 \cite{Perspectives} and 
its finite-T extension KST2 \cite{KarasievSjostromTrickey12A} 
do predict binding, but both 
also have worse numerical convergence than VT84F because of the 
same wrong large-$s$ limit.

\subsection{Ab initio molecular dynamics: scaling}

We did OF-DFT AIMD and KS AIMD simulations for  H 
again at material density 0.983 g/cm$^3$ ($r_s =1.40$ bohr) with 128 atoms 
in the simulation cell
using the NVT ensemble regulated by the Andersen thermostat.  Maximum 
temperatures were 4,000,000 K and 181,000 K for OF-DFT and KS MD
respectively.  Some of the physical results ({\it e.g.} equation of state,
pair-correlation functions, etc.) have been published \cite{VT84F} and
others will be published systematically elsewhere.  Here the focus
is on comparative computational performance. 

\begin{figure}
\includegraphics*[angle=-00,width=8.0cm]{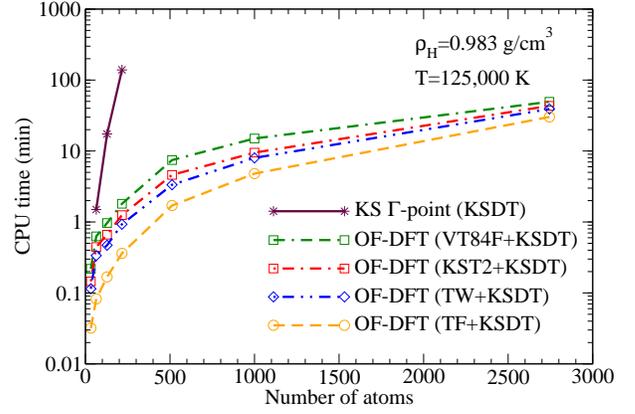}
\caption{
CPU time per MD step for OF-DFT calculations with VT84F, KST2, TF, and TW 
non-interacting free-energy functionals and PZ XC for H 
at material density $\rho_{\rm H}=0.983$ g/cm$^3$ and  ${\mathrm T}=125,000$ K
with 32, 64, 128, 216, 512, 1000, and 2744 atoms in the simulation cell.
Corresponding data for KS calculations with 64, 128 and 216 H atoms 
in the simulation cell ($\Gamma$-point) are shown for comparison.
All calculations used a single core of an Intel $i5$ 3.2GHz CPU.
}
\label{time-vs-N}
\end{figure}

\begin{figure}
\includegraphics*[angle=-00,width=8.0cm]{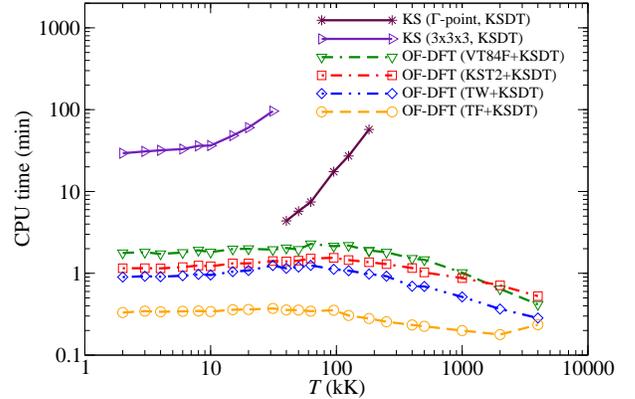}
\caption{
CPU time per MD step as a function of temperature for KS and OF-DFT 
calculations with VT84F, KST2, TF, and TW 
non-interacting free-energy functionals and PZ XC for H 
at material density $\rho_{\rm H}=0.983$ g/cm$^3$
with 128 atoms in the simulation cell.
All calculations used a single core of an Intel $i5$ 3.2GHz CPU.
}
\label{time-vs-T}
\end{figure}

Figure \ref{time-vs-N} compares OF-DFT and KS CPU time per MD step 
for hydrogen at ${\mathrm T}=125,000$ K
as a function of the number of atoms $N$ in the simulation cell.
All the timings are on a single core so as not to penalize the KS
calculations for their serial overhead.    
OF-DFT AIMD exhibits almost linear scaling $O(N\ln N)$ with system size
\cite{CPL.Carter.2009}, with 45 minutes of CPU time per MD step 
for $N=2744$. KS AIMD exhibits the expected approximately $N_b^3$ scaling, 
($N_b>N$ is the number of thermally occupied one electron states).  In
the university context, that scaling makes such 
simulations prohibitively expensive for $N>216$ atoms. 

CPU time as a function of temperature (still for $N = 128$ hydrogen atoms in
at material density 0.983 g/cm$^3$) is shown in Figure \ref{time-vs-T}. 
OF-DFT times vary slightly because of different convergence rates 
at different temperatures. 
For extremely high T, the CPU time decreases as the system
tends toward the classical high-T limit.  The  
KS CPU time per MD step in contrast increases drastically with increasing
T because of the involvement of a growing number of partially
occupied one-electron states $N_b$.  Again, in our academic context,
this temperature scaling makes KS AIMD for
the given material density prohibitively expensive for T$>200,000$ K.

\subsection{Recommended functionals}

Detailed accuracy comparisons among different orbital-free functionals and 
comparison between OF-DFT and reference KS results are outside the 
scope of the present work.
Previous publications
\cite{Perspectives,PRB80,KarasievTrickeyCPC2012,KarasievSjostromTrickey12A,VT84F} 
provide relevant comparisons. For guidance in using the new 
interface and coding presented here, the 
main conclusions from those studies can be summarized as follows.
(1) The standard non-interacting GGA functionals,  including the empirical 
PBETW, and non-empirical APBEK ones, do not provide a qualitatively 
correct treatment of binding 
in simple molecules and in solids such as sc-H and fcc-Al. 
(2) In contrast, both the VT84F and PBE2/KST2 
non-interacting functionals 
conversely provide at least semi-quantitatively correct predictions at 
all temperatures and have proper T=0 K limits.  (3) 
The temperature dependence of the XC free energy becomes essential at elevated
T. (4)  The non-empirical VT84F functional provides almost identical 
results with those from the mildly empirical PBE2/KST2 and has the added 
benefit of better SCF convergence because of its 
correct large-$s$ behavior.  On the basis of these facts, our 
current recommendation for OF-DFT AIMD calculations is to use
the VT84F non-interacting free energy in combination with
the KSDT XC free energy.  \vspace*{8pt}

\section{Conclusions}

We have described the essential ingredients and modifications in 
our implementation of free energy functionals
in the orbital-free {\sc Profess} code.  And we have described the
interfacing of that code with the {\sc Quantum-Espresso} code
to provide a consistent suite with which to do both OF-DFT AIMD and
KS AIMD calculations at all temperatures.   Non-interacting
free-energy one-point functionals defined within the finite-T
generalized gradient approximation provide an adequate  
quantum statistical mechanical description of the electrons, thereby
reducing the computational cost of using the non-local two-point 
(ground state) functionals in the original {\sc Profess} code.
Our {\sc Profess@Q-Espresso} interface and all patches 
to {\sc Profess 2.0} and {\sc Quantum Espresso} 5.0.3 are available
by download from http://www.qtp.ufl.edu/ofdft and by request to
the authors. 

\section{Acknowledgments}
VVK and SBT  were supported by the U.S.\ Dept.\ of Energy TMS program,  
grant DE-SC0002139, as was the initial part of the effort by TS.  The
latter part of his work was supported by the Office of Fusion Energy
Sciences of DOE.    
We acknowledge with thanks the provision of computational resources
and technical support by the University of Florida High-Performance 
Computing Center.

\end{document}